\documentclass[]{JHEP3}%
\usepackage{graphics}      
\usepackage{epsfig,latexsym}

\newcommand{\be}{\begin{equation}}
\newcommand{\ee}{\end{equation}}
\newcommand{\bea}{\begin{eqnarray}}
\newcommand{\eea}{\end{eqnarray}}
\newcommand{\lsim}{\mbox{\raisebox{-.6ex}{~$\stackrel{<}{\sim}$~}}}
\newcommand{\gsim}{\mbox{\raisebox{-.6ex}{~$\stackrel{>}{\sim}$~}}}

\def\Fbox{F\left(\frac{\Box}{m_s^2}\right)}

\def\done{\delta^{(1)}}
\def\dtwo{\delta^{(2)}}

\title{Large Nongaussianity from Nonlocal Inflation}

\author{N.\ Barnaby, J.\ M.\ Cline\\
Physics Department, McGill University, Montr\'eal, Qu\'ebec, Canada H3A
2T8\\
E-mail: \email{barnaby@physics.mcgill.ca}, 
\email{jcline@physics.mcgill.ca}}

\preprint{}
\abstract{
We study the possibility of obtaining large nongaussian signatures in the Cosmic
Microwave Background in a general class of single-field nonlocal hill-top inflation models.  
We estimate the nonlinearity parameter $f_{NL}$ which characterizes nongaussianity
in such models and show that large nongaussianity is possible.  
For the recently proposed $p$-adic inflation model we find that $f_{NL} \sim 120$ 
when the string coupling is order unity.  We show that large nongaussianity is 
also possible in a toy model with an action similar to those which arise in string field theory.
}

\begin{document}

\section{Introduction}

Though the simplest models of inflation yield a negligible degree of nongaussianity
in the Cosmic Microwave Background (CMB) there has been considerable interest recently in 
constructing models which \emph{can} give
rise to large nongaussian signatures \cite{BMR}-\cite{post_inf_NG2} (see \cite{NGreview} for
a review).  Nongaussianity is typically characterized by the dimensionless nonlinearity 
parameter $f_{NL}$ which is of the order of $|n_s-1| \ll 1$\footnote{Nongaussianity 
can further
be characterized using the trispectrum, which is also small in the simplest models 
\cite{trispectrum}.}
(where $n_s$ is the spectral index) in conventional models \cite{BMR}-\cite{SeeryLidsey}.  
The current observational limit is $|f_{NL}|\lsim 300$ \cite{limit} (for the WMAP 3 year data 
\cite{WMAP3}) and future missions are expected to be able to probe $f_{NL}$ as small as order 
unity \cite{future}.  Though it has been difficult to find models which can yield 
large nongaussianity (indeed, observation of $|f_{NL}| \gsim 5$ would be strongly indicative
of some novelty in the dynamics driving inflation) there exist several examples in the 
literature:
\begin{enumerate}
  \item Single field inflationary models in which the inflaton has a small sound speed 
        \cite{large_nongauss}, as in \cite{DBI}.
  \item Single field models in which the inflaton potential has a sharp feature \cite{feature}.
  \item Hybrid inflation \cite{hybrid} can generate significant nongaussianity during the 
        preheating phase for certain values of the model parameters \cite{BC1,BC2}.  
        Nongaussianity from preheating has also been considered in \cite{preheatNG}.
  \item The curvaton model \cite{curvaton}.
  \item Ghost inflation \cite{ghost_inflation}.
\end{enumerate}
The simplest multi-field models do not yield large nongaussianity \cite{multi_field},
however, this need not be true also of more complicated models.
In this paper we show that large nongaussianity may also be generated in certain inflationary
models based on nonlocal field theory.  In particular, we will show that large nongaussianity
is possible in the recently proposed $p$-adic inflation model \cite{pi}.  

Recently, cosmological applications of field theories containing an infinite number of 
derivatives have attracted considerable interest in the literature 
\cite{pi}-\cite{phantom}.  In \cite{tirtho} higher derivative modifications of gravity 
were considered and it was shown that such theories can have novel features such as improved
ultraviolet (UV) behaviour and the existence of nonsingular bouncing solutions.  In 
\cite{justin} it was shown that similar nonlocal gravity theories can lead to self-inflation.  

In this paper we will be interested in the cosmology of scalar field theories containing an 
infinite number of derivatives (we assume a standard gravitational sector) as in \cite{lidsey} 
and \cite{gen_nonlocal1}-\cite{phantom}.  Nonlocal scalar field theories of this type can be 
derived from string theory, the most popular examples being the $p$-adic string theory
\cite{witten}-\cite{padic} and cubic string field theory \cite{SFT_action}-\cite{SFT}.  
In both cases the scalar field is a tachyon describing the instability of some unstable 
D-brane configuration.  Our interest in these types of nonlocal field theories is motivated
by their novel cosmological behaviour.  Nonlocal field theories can exhibit a variety of 
interesting cosmological properties including the possibility of slow roll inflation with steep 
potentials 
\cite{pi,lidsey}, bouncing cosmologies \cite{gen_nonlocal2} and an equation 
of state $w < -1$ for dark energy within a sensible microscopic theory 
\cite{gen_nonlocal1,gen_nonlocal2,phantom}.

In this paper we will focus on nonlocal field theories in which inflation is realized near the 
top of an unstable maximum of the potential (we refer to such constructions as nonlocal 
hill-top \cite{hilltop} inflation models).  In \cite{pi} a particular model based on $p$-adic 
string theory, 
$p$-adic inflation, was proposed.  A novel feature was the possibility of slow roll dynamics for 
the inflaton, despite the presence of a steep potential, which suggests an intriguing
possible resolution for the fine tuning problems which typically plague inflationary model
building.  In \cite{lidsey} a more general 
analysis of nonlocal hill-top inflation was performed and it was shown that this behaviour 
is possible also in other models.  In this paper we study the possibility of obtaining 
nongaussian perturbations in general models of the form considered in \cite{lidsey}, specializing 
at the end to the case of $p$-adic inflation.  We find that in $p$-adic inflation 
$f_{NL} \sim 120$ is possible for $g_s \sim 1$.  We show that large nongaussianity is possible
also in other models.

The plan of the paper is as follows.  In section \ref{sec:hilltop} we consider
the inflationary dynamics of  nonlocal hill-top inflation, generalizing
the results of \cite{lidsey} to include the possibility of a $\phi^3$ term in the 
scalar field potential.  In section \ref{p_compare} we show how these results apply to 
the case of $p$-adic inflation.  In section \ref{sec:ng} we provide
an estimate for the nonlinearity parameter in nonlocal hill-top inflation.  In section 
\ref{sec:png} we use this result to estimate the nonlinearity parameter in $p$-adic inflation,
 showing that $|f_{NL}| \gg 1$ is possible.  Finally, we consider a toy model 
with an exponential kinetic function (which is typical of string field theory Lagrangians)
showing that large nongaussianity is possible.

\section{Nonlocal Hill-top Inflation}
\label{sec:hilltop}

We consider general nonlocal theories of the form \cite{lidsey} 
(see also \cite{gen_nonlocal1,gen_nonlocal2})
\begin{equation}
\label{action}
  \mathcal{L} = \gamma^4\left[\frac{1}{2}\phi \,\Fbox \phi - U(\phi) \right]
\end{equation}
where we assume a hill-top potential of the form
\begin{equation}
\label{pot}
  U(\phi) = U_0 - \frac{\mu^2}{2}\phi^2 + \frac{g}{3}\phi^3 + \cdots
\end{equation}
where the $\cdots$ denotes terms of order $\mathcal{O}(\phi^4)$ and higher.  Such terms
will generically be present, but since we are interested in the dynamics close to the
false vacuum $\phi = 0$, they will play a subleading role in our calculation.  Indeed,
the potential (\ref{pot}) should only be thought of as an effective description close to
$\phi = 0$ since it must be supplemented by additional terms to ensure that $U(\phi)$
is bounded from below---at least on one side.  We use metric signature $\eta_{\mu\nu} = \mathrm{diag}(-1,+1,+1,+1)$ so 
that $\Box = -\partial_t^2 + \nabla^2$ in flat space.
In the above we take $\phi$, $U_0$, $\mu$ and $g$ to be dimensionless
while $\gamma$, $m_s$ have mass dimension one.  With no loss of generality we 
can set $F(0) = 0$, since $F(0)$ can always be absorbed into 
the definition of $\mu$.  The differential operator $\Fbox$ should be understood
as a series expansion
\begin{equation}
\label{Fseries}
  \Fbox = \sum_{n=0}^{\infty} c_n\left(\frac{\Box}{m_s^2}\right)^{n}
\end{equation}
where 
\[
c_n = \frac{1}{n!}\left.\frac{d^{(n)}F(z)}{dz^{(n)}}\right|_{z=0}
\]
and $c_0 = 0$.  Lagrangians of the form (\ref{action}) arise frequently
in string field theory, in which case $m_s$ coincides with the string mass scale.  
In the following we keep $m_s$ unspecified.

Examples of theories of the form (\ref{action}) include the $p$-adic string theory
\cite{witten} (see also \cite{zweibach,padic})
\begin{equation}
\label{p_action}
  \mathcal{L} = \frac{m_s^4}{g_p^2}\left[-\frac{1}{2}\psi p^{-\frac{\Box}{m_s^2}} \psi 
                + \frac{1}{p+1}\psi^{p+1}\right]
\end{equation}
with
\[
  \frac{1}{g_p^2} = \frac{1}{g_s^2}\frac{p^2}{p-1}
\]
and, in order to put (\ref{p_action}) in the form (\ref{action}), we take
$\psi = \phi + 1$. In (\ref{p_action}) $m_s$ and $g_s$ are the string mass and coupling
respectively.  The action (\ref{p_action}) was derived assuming $p$ to be a prime number,
though it appears to make sense for any integer value of $p$.
A second popular example is the tachyon action \cite{SFT_action} (see also
\cite{gen_nonlocal1,SFT_action2,SFT})
\begin{equation}
\label{sft_action}
  \mathcal{L} = \frac{m_s^4}{g_s^2}\left[ 
  \frac{1}{2}\psi\left(1 + \lambda^2 \frac{\Box}{m_s^2}\right)e^{-\frac{\Box}{4 m_s^2}}\psi
   - \frac{1}{4}\psi^4 \right]
\end{equation}
where $\lambda^2 \cong 0.9556$.  This tachyon action is derived in cubic superstring field theory
(SFT) in a flat background, incorporating only massless fields.

\subsection{Background Dynamics}

We first consider the homogeneous dynamics of the theory (\ref{action}).
The Klein-Gordon equation for the scalar field $\phi$ is
\begin{equation}
\label{KGbkg}
  \Fbox \phi(t) = -\mu^2 \phi(t) + g \phi^2(t)
\end{equation}
The Friedmann equation is
\begin{equation}
\label{friedmann}
  3 H^2 = \frac{1}{M_p^2} T_{00}
\end{equation}
where the stress tensor is given by \cite{lidsey} (see also \cite{stress})
\begin{eqnarray}
  \frac{1}{\gamma^4} T_{\mu\nu} &=& \sum_{l=1}^{\infty}c_l\, m_s^{-2l}\sum_{j=0}^{l-1}
  \left[ \left(\partial_\mu \Box^{j} \phi \right)\left(\partial_\nu \Box^{l-1-j}\phi\right)
   - \frac{1}{2}g_{\mu\nu}
  \left(\partial_\alpha \Box^{j} \phi \right)\left(\partial_\alpha \Box^{l-1-j}\phi\right)
    \right. \nonumber \\
  &-& \left. \frac{1}{2} g_{\mu\nu} \left(\Box^{j}\phi\right)\left(\Box^{l-j}\phi\right)\right]
  + \frac{1}{2}\left[\phi \sum_{l=0}^{\infty}c_l \left(\frac{\Box}{m_s^2}\right)^{l} \phi 
  - 2 V\right] \label{Tmunu}
\end{eqnarray}
In order to solve the system (\ref{KGbkg},\ref{friedmann}) we make the ansatz of
a series expansion
\begin{eqnarray}
  \phi(t) &=& \sum_{r=1}^{\infty} \phi_r e^{r \lambda t} \nonumber \\
  H(t) &=& H_0 - \sum_{r=1}^{\infty} \phi_r e^{r \lambda t} \label{series_ansatz}
\end{eqnarray}
and solve for $\lambda$ and $\{\phi_r$, $H_r\}$ order-by-order in powers of $u = e^{\lambda t}$.
We have parameterized the solution so that as $t \rightarrow -\infty$ the field $\phi$ sits
at the unstable maximum $\phi = 0$ and the universe undergoes de Sitter expansion with Hubble
constant $H_0$.  We use the freedom to choose the origin of time to set $\phi_1 \equiv 1$.
We also set $H_1 \equiv 0$ which is consistent because $T_{\mu\nu}$ does not contain any terms
 linear in $\phi$.  Thus, to solve the equations of motion up to order 
$\mathcal{O}(u^2)$ the truncated expansion
\begin{eqnarray}
  \phi(t) &=& u + \phi_2 u^2 \label{trunc_series_phi} \\
  H(t) &=& H_0 - H_2 u^2 \label{trunc_series}
\end{eqnarray}
suffices.

\subsubsection{Scalar Field Evolution}

We now proceed to solve (\ref{KGbkg}) using (\ref{trunc_series_phi}).  It was shown in
\cite{pi} that
\begin{equation}
\label{box_phi}
  (-\Box)^{n}\phi = +\left(\lambda^2 + 3H_0\lambda\right)^{n} u 
  + \left(4\lambda^2 + 6H_0\lambda\right)^{n} \phi_2 u^2 + \mathcal{O}(u^3)
\end{equation}
so that the left-hand-side of (\ref{KGbkg}) is
\begin{equation}
\label{lhs}
  \Fbox \phi = F\left(-\frac{\lambda^2 + 3H_0\lambda}{m_s^2}\right) u
  + F\left(-\frac{4\lambda^2 + 6H_0\lambda}{m_s^2}\right) \phi_2 u^2
  + \mathcal{O}(u^3)
\end{equation}
while the right-hand-side of (\ref{KGbkg}) gives
\begin{equation}
\label{rhs}
  -\mu^2 \phi + g \phi^2 = -\mu^2 u + (g-\mu^2 \phi_2) u^2 + \mathcal{O}(u^3)
\end{equation}
Matching at linear order in $u$ implies
\begin{equation}
\label{first}
  F\left(-\frac{\lambda^2 + 3H_0\lambda}{m_s^2}\right) = -\mu^2
\end{equation}
It is convenient to define
\begin{equation}
\label{omega}
  \omega^2 = -m_s^2 F^{-1}\left(-\mu^2\right)
\end{equation}
since consistency of this approach requires that $F^{-1}$ exists and is single-valued, 
so that (\ref{first}) takes the form
\begin{equation}
\label{characteristic}
  \lambda^2 + 3 H_0 \lambda - \omega^2 = 0
\end{equation}
It is straightforward to solve (\ref{characteristic}) for $\lambda$
\[
  \lambda = -\frac{3H_0}{2} \pm \frac{3H_0}{2}\sqrt{1+\frac{4\omega^2}{9H_0^2}}
\]
The slow-roll solution corresponds to taking the positive root and assuming that
$\omega^2 \ll H_0^2$ so that
\begin{equation}
\label{lambda}
  \lambda \cong \frac{\omega^2}{3 H_0}
\end{equation}
We also define a dimensionless parameter 
\begin{equation}
\label{eta}
  \eta = -\frac{\omega^2}{3 H_0^2}
\end{equation}
so that slow roll corresponds to $|\eta| \ll 1$ (notice that $\eta < 0$ in this model).  
For the slowly rolling solution we have $\lambda / H_0 = |\eta| \ll 1$
and $\lambda^2 \ll 3 H_0 \lambda$ so that the evolution is friction-dominated in the usual
sense, namely $\ddot{\phi} \ll H \dot{\phi}$ when $|\eta| \ll 1$.  
In passing, notice that to leading order in the small-$u$ expansion the field
obeys the eigenvalue equation
\begin{equation}
\label{evalue_bkg}
  \Box \phi = -\omega^2 \phi
\end{equation}
However, the correspondence between the solutions of (\ref{evalue_bkg}) and the solutions
of (\ref{KGbkg}) breaks down beyond leading order in the small-$u$ expansion.

We now solve (\ref{KGbkg}) up to second order in the small-$u$ expansion.  
Notice that the arguments of $F$ in (\ref{lhs}) are approximately $-\omega^2 / m_s^2$
and $-2\omega^2 / m_s^2$ for the coefficients of the $\mathcal{O}(u)$ and 
$\mathcal{O}(u^2)$ terms respectively.  Matching (\ref{lhs}) and (\ref{rhs}) at order $u^2$ 
gives
\begin{equation}
\label{second}
  \phi_2 \cong \frac{g}{\mu^2 + F\left(-2\frac{\omega^2}{m_s^2}\right)}
\end{equation}
for $|\eta| \ll 1$.
Notice that in the case $g=0$ we have $\phi_2 = 0$ and the correspondence between the
solutions of (\ref{evalue_bkg}) and the solutions of (\ref{KGbkg}) holds quite generally.
As we have pointed out previously, however, for $g \not= 0$ this equivalence breaks down
at order $u^2$.

\subsubsection{Friedmann Equation}

It now remains to determine the coefficients $H_0$, $H_2$ in (\ref{trunc_series}).  In order
to determine $H(t)$ up to order $u^2$ we need only consider $\phi(t)$ up to first order 
in the small-$u$ expansion.  This is so because $T_{\mu\nu}$ does not contain any term
linear in $\phi$ (see eq.\ (\ref{Tmunu})).  Working strictly with $\phi(t) = u$ the scalar 
field obeys (\ref{evalue_bkg}) so that it is straightforward to resum the infinite series in 
(\ref{Tmunu}).  The result is
\begin{equation}
\label{T00}
  \gamma^{-4}\, T_{00} = U_0 - \frac{\omega^2}{2m_s^2}F'\left(-\frac{\omega^2}{m_s^2}\right)
  \left[1 - \frac{|\eta|}{3}\right] u^2 + \mathcal{O}(u^3)
\end{equation}
Then we have
\begin{equation}
\label{T00_sr}
  \frac{\sqrt{T_{00}}}{\sqrt{3}\,M_p} 
  \cong \frac{\gamma^2 U_0^{1/2}}{\sqrt{3}\, M_p}
  \left[ 1 - \frac{\omega^2}{4 m_s^2 U_0}F'\left(-\frac{\omega^2}{m_s^2}\right) u^2 
  + \mathcal{O}(u^3)   \right]
\end{equation}
at leading order in the $\eta$ slow roll parameter.
The fact that there is no $\mathcal{O}(u)$ term on the right-hand-side of (\ref{T00})
demonstrates that it was consistent to set $H_1 = 0$ in (\ref{trunc_series}).
From the Friedmann equation
\[
  H(t) = \frac{\sqrt{T_{00}}}{\sqrt{3}\,M_p} 
\]
it is straightforward to identify $H_0$, $H_2$ by equating (\ref{trunc_series}) to 
(\ref{T00_sr}).  We find that
\begin{equation}
\label{H0}
  H_0 = \frac{\gamma^2 U_0^{1/2}}{\sqrt{3}\, M_p}
\end{equation}
and
\begin{equation}
\label{H2}
  \frac{H_2}{H_0} = \frac{\omega^2}{4 m_s^2 U_0}F'\left(-\frac{\omega^2}{m_s^2}\right)
\end{equation}
Or, using (\ref{omega}) to eliminate the derived parameter $\omega^2$, we have
\begin{equation}
\label{H2_simple}
  \frac{H_2}{H_0} = -\frac{1}{4 U_0} F^{-1}(-\mu^2) F'\left[F^{-1}(-\mu^2)\right]
\end{equation}

It is useful to define a second dimensionless ``slow roll'' parameter by
\begin{equation}
\label{epsilon}
  \epsilon(t) = \frac{H_2}{H_0} e^{2\lambda t}
\end{equation}
It is clear that once $\epsilon(t) = 1$ then the perturbative expansion (\ref{trunc_series})
breaks down and our solution is no longer valid.  At this point one may
assume that inflation has ended 
and hence the condition $\epsilon(t_{\mathrm{end}}) = 1$
implicitly defines the time $t_{\mathrm{end}}$ at which inflation ends.  It is worth
pointing out that the definition (\ref{epsilon}) differs from the usual $\epsilon$ slow
roll parameter since
\[
  -\frac{\dot{H}}{H^2} \cong 2 |\eta| \epsilon
\]
with our definitions (\ref{eta}, \ref{epsilon}).  A quantity in which we will be interested
in later on is the value of $u = e^{\lambda t}$ at the end of inflation.  Using 
(\ref{H2_simple}) we see that
\begin{equation}
\label{u_end}
  \frac{1}{u_{\mathrm{end}}^2} = -\frac{1}{4 U_0} F^{-1}(-\mu^2) F'\left[F^{-1}(-\mu^2)\right]
\end{equation}

\subsection{Fluctuations}

We now study the scalar field fluctuations about a homogeneous background,
\begin{equation}
\label{pert}
  \phi(t,{\bf x}) = \phi_0(t) + \delta \phi(t,{\bf x})
\end{equation}
As we have shown the background solution during inflation is described by
$\phi_0(t) \cong e^{\lambda t}$, $H(t) \cong H_0- H_2 e^{2\lambda t}$.  
In order to solve for the fluctuation $\delta \phi$ we treat the background as pure de 
Sitter space (which is equivalent to working to zeroth order in the small-$u$ expansion so that
$\phi_0 = 0$ and $H = H_0$).  In this limit the fluctuation $\delta \phi$ obeys the equation
\begin{equation}
\label{fluct}
  \Fbox \delta \phi = -\mu^2 \delta \phi
\end{equation}
We can obtain solutions of (\ref{fluct}) by choosing $\delta\phi$ to satisfy the eigenvalue
equation
\begin{equation}
\label{evalue}
  \Box \delta \phi = -\omega^2 \delta \phi
\end{equation}
with $\omega^2$ given by (\ref{omega}) as above.
The solutions of (\ref{evalue}) are well known.  However, to make contact with the usual
treatment of cosmological perturbations we need to define a field in terms of which the action
looks canonical.  This presents a difficulty because, in general, there is no local
field redefinition which will bring the kinetic term $\phi \Fbox \phi$ into canonical form.
However, as in \cite{pi,lidsey}, we can circumvent this difficulty by noticing that 
(\ref{fluct}) can be derived from the perturbed Lagrangian
\begin{equation}
\label{L_pert}
  \mathcal{L}^{(2)} = \frac{\gamma^4}{2}\left[\delta\phi\Fbox\delta\phi 
  + \mu^2 (\delta\phi)^2\right]
\end{equation}
Then, for on-shell fields (that is, when (\ref{evalue}) is satisfied),
\begin{eqnarray*}
  \mathcal{L}^{(2)}_{\mathrm{on-shell}} &=& 
  \frac{\gamma^4}{2}\left[\delta\phi F\left(-\frac{\omega^2}{m_s^2}\right)\delta\phi 
  + \mu^2 (\delta\phi)^2\right] \\
  &=& \frac{\gamma^4}{2}\left[
  \delta\phi \frac{1}{(-\omega^2)}F\left(-\frac{\omega^2}{m_s^2}\right)(-\omega^2)\delta\phi 
  + \mu^2 (\delta\phi)^2\right] \\
  &=& \frac{\gamma^4}{2}\left[
  \delta\phi \frac{1}{(-\omega^2)}F\left(-\frac{\omega^2}{m_s^2}\right)\Box \delta\phi 
  + \mu^2 (\delta\phi)^2\right] \\
  &=& \frac{1}{2}\delta \varphi \Box \delta \varphi + \frac{\omega^2}{2}(\delta \varphi)^2
\end{eqnarray*}
where we have defined
\begin{equation}
\label{varphi}
  \delta \varphi = A \delta\phi
\end{equation}
with
\begin{equation}
\label{A}
  A^2 = \gamma^4 \frac{1}{(-\omega^2)}F\left(-\frac{\omega^2}{m_s^2}\right)
      = \frac{\gamma^4}{m_s^2} \frac{-\mu^2}{F^{-1}(-\mu^2)}
\end{equation}
In the second equality in (\ref{A}) we have used (\ref{omega}) to eliminate the derived parameter
$\omega$.  Thus $\delta\varphi$ is the variable which has canonical kinetic term in the 
action.\footnote{Our choice of normalization coincides with the definition of a ``canonical''
inflaton advocated in \cite{pi}.  However, this choice differs from the 
normalization employed in \cite{lidsey},where the inflaton was normalized
in such a way that the stress tensor $T_{\mu\nu}$ takes canonical form, though the 
kinetic term in the action does not.  In the case of $p$-adic inflation,
our main interest, the discrepancy is only a factor of $\sqrt{\ln p}$ which is of order
unity for the values of $p$ which we consider.}

Notice that the definition of the canonical field, eq.\ (\ref{varphi}), has been derived by
studying the linearized theory.  However, we will continue to adopt this definition even up
to second order in cosmological perturbation theory when we compute the nongaussianity
in section \ref{sec:ng}.  This is justified since in the ADM formalism 
\cite{ADM} it suffices to use
the free-field solution to compute the interaction Hamiltonian.  This is so because the
terms in the Lagrangian which would provide nonlinear corrections to the free-field dynamics
are always multiplied by terms proportional to first order equations of motion, and
hence these corrections vanish \cite{large_nongauss}.  
Similar comments apply to our construction of the 
curvature perturbation in subsection \ref{curv_sub_sect}.
In any case, contributions to
$f_{NL}$ coming from the free-field dynamics must always be present and, barring anomalous 
cancellation, the contribution of such terms provides a lower bound on the
actual nongaussianities produced.

We now proceed to solve (\ref{evalue}), bearing in mind that $\delta \varphi$ is the appropriate
canonically normalized field.  We write the Fourier transform of the inflaton fluctuation
as
\begin{equation}
\label{Fourier}
  \delta \varphi(t,\vec{x}) = \int \frac{d^{\,3}k}{(2\pi)^{3/2}} e^{i{\bf k}\cdot{\bf x}}
  \xi_{\bf k}(t)
\end{equation}
where the operator-valued Fourier coefficients $\xi_k(t)$ can be decomposed into
annihilation/creation operators $a_k,a_k^\dagger$ and c-number-valued mode functions $\varphi_k$
as
\begin{equation}
\label{xi}
  \xi_{\bf k}(t) = a_{\bf k} \varphi_{\bf k}(t) 
  + a_{\bf -k}^{\dagger} \varphi_{\bf -k}^{\star}(t) 
\end{equation}
The mode functions $\varphi_k(t)$ are given by
\begin{equation}
\label{mode}
  \varphi_k(t) = \frac{1}{2}\sqrt{\frac{\pi}{a^3 H_0}}\,e^{\frac{i\pi}{2}(\nu+1/2)}
            \, H_\nu^{(1)}\!\left(\frac{k}{a H_0}\right)
\end{equation}
where the order of the Hankel functions is
\begin{equation}
\label{nu}
  \nu = \sqrt{\frac{9}{4} + \frac{\omega^2}{H_0^2}}
\end{equation}
and of course $a = e^{H_0 t}$. 
In writing (\ref{mode}) we have used the usual Bunch-Davies
vacuum normalization so that on small scales, $k \gg a H_0$, one has
\[
  |\varphi_k| \cong \frac{a^{-1}}{\sqrt{2k}}
\]
which reproduces the standard Minkowski space fluctuations.  This is the usual procedure in
cosmological perturbation theory.  However, we note that the quantization of the theory
(\ref{action}) is not transparent and it  might turn out that the usual
prescription is incorrect in the present context.  We defer this and other subtleties to future
investigation.  On large scales, $k \ll a H_0$, the solutions (\ref{mode}) behave as
\[
  |\varphi_k| \cong \frac{H}{\sqrt{2k^3}}\left(\frac{k}{a H_0}\right)^{3/2-\nu}
\]
which gives a large-scale power spectrum for the fluctuations
\[
  P_{\delta\varphi} = \left(\frac{H_0}{2\pi}\right)^2\left(\frac{k}{a H_0}\right)^{n_s-1}
\]
with spectral index
\begin{equation}
\label{n}
  n_s - 1 = 3 - 2\nu \cong 2 \eta
\end{equation}
where $\eta$ is given by (\ref{eta}).  Since $\eta < 0$ this model always gives
a red-tilted spectrum, in agreement with the latest WMAP data \cite{WMAP3}.

Using (\ref{eta}) and (\ref{H0}) we can  rewrite (\ref{n}) as
\begin{equation}
\label{gamma_eliminate}
  \frac{1}{\gamma^4} = \frac{|n_s-1|}{2}\frac{U_0}{M_p^2 m_s^2}\frac{1}{|F^{-1}(-\mu^2)|}
\end{equation}
which can be use to eliminate $\gamma$ in favour of other parameters.

It is worth noting that the equivalence between the solutions of (\ref{evalue}) and 
(\ref{fluct}), which implies the equivalence between local and nonlocal theories described
in \cite{lidsey}, appears only at linear order in perturbation theory.  Beyond linear
order (as long as $g\not= 0$) this equivalence breaks down, as we illustrate for the case
of $p$-adic inflation in appendix A.  The breakdown of this equivalence, once 
interactions are included, is crucial for understanding why nonlocal theories of the type 
(\ref{action}) can give rise to significant nongaussianity. Indeed, if it were true
that the equivalence persisted at all orders in
perturbation theory then the calculation of $f_{NL}$ in the theory (\ref{action})
would be exactly equivalent to a calculation of $f_{NL}$ in some local field theory
where $f_{NL} \ll 1$ is generic.

\subsection{The Curvature Perturbation}
\label{curv_sub_sect}

A full calculation of the power spectrum should include metric perturbations and also
deviation of the background expansion from pure de Sitter.  Such a computation is beyond
the scope of the present paper.  Here we neglect metric fluctuations and take $H \cong H_0$
to compute the field perturbation, as we have done in the last subsection.  (These 
approximations reproduce the full calculation up to acceptable accuracy in the standard
theory.)  We assume that the curvature perturbation is given by
\begin{equation}
\label{zeta}
  \zeta = -\frac{H}{\dot{\varphi}_0}\delta \varphi
\end{equation}
where $\varphi_0 = A\phi_0 = A u$.  To evaluate the prefactor $H / \dot{\varphi}_0$
we must work beyond zeroth order in the small-$u$ expansion.  We take $\phi_0 = u$
to compute $H / \dot{\varphi}_0$ even though the perturbation $\delta \varphi$ is computed in 
the limit that $\phi_0 = 0$.  This should reproduce the full answer up to $\mathcal{O}(u)$ 
corrections.  It will come in 
handy later on to define $\zeta = c_{\zeta} \delta \varphi$ so that the prefactor $c_{\zeta}$ is
\[
  c_{\zeta} = -\frac{H}{\dot{\varphi}_0} = -\frac{H_0}{A \lambda}\frac{1}{u}
\]
To compute the spectrum this must be evaluated at horizon crossing, $N_e$ e-foldings
before the end of inflation.  Noting that
\[
  u(t) = a(t)^{|\eta|}
\]
and the scale factor at horizon crossing is related to that at the end of inflation by
$a_{\star} = e^{-N_e}a_{\mathrm{end}}$,  it follows that 
\begin{equation}
\label{ustar}
 u_\star = e^{-N_e|\eta|} u_{\mathrm{end}} = e^{-\frac{N_e}{2}|n_s-1|}u_{\mathrm{end}}
\end{equation}
while $ u_{\mathrm{end}}$ is given by (\ref{u_end}).  In the above 
$u_\star = u(t_\star) = e^{\lambda t_\star}$ is the value of $u(t)$ at the time
of horizon crossing, $t_\star$.  Further, since $\lambda/H_0 = |\eta| = |n_s-1|/2$ and 
using (\ref{A}, \ref{u_end}, \ref{ustar}) we can write the prefactor 
$c_{\zeta} = -H / \dot{\varphi}_0$ at the time of horizon crossing as
\begin{eqnarray}
  c_{\zeta}^2 = 
  \left( -\frac{H}{\dot{\varphi}_0}\right)^2 &=& \frac{4}{|n_s-1|^2}\frac{m_s^2}{\gamma^4}
  \frac{F^{-1}(-\mu^2)}{(-\mu^2)} \frac{e^{N_e |n_s-1|}}{u_{\mathrm{end}}^2} \nonumber \\
  &=& \frac{e^{N_e |n_s-1|}}{|n_s-1|^2}\frac{m_s^2}{\gamma^4 U_0}\frac{1}{\mu^2}
  \left[F^{-1}(-\mu^2)\right]^2F'\left[F^{-1}(-\mu^2)\right]
  \label{prefactor}
\end{eqnarray}
it follows that the power spectrum of the curvature perturbation is
\begin{equation}
\label{pwr}
  P_{\zeta} = A_{\zeta}^2 \left(\frac{k}{a H_0}\right)^{n_s-1}
\end{equation}
where
\begin{equation}
\label{Azeta}
  A_\zeta^2 = \left( -\frac{H}{\dot{\varphi}_0}\right)^2\left(\frac{H_0}{2\pi}\right)^2
  = \ c_{\zeta}^2 \frac{H_0^2}{(2\pi)^2} \ \cong\  25\times 10^{-10}
\end{equation}
Using (\ref{prefactor}) and (\ref{H0}) we arrive at the result
\begin{equation}
\label{amp}
  A_{\zeta}^2 = \frac{1}{12\pi^2}\left(\frac{m_s}{M_p}\right)^2\frac{e^{N_e|n_s-1|}}{|n_s-1|^2}
 \, G(\mu^2)
\end{equation}
where we have defined the function
\begin{equation}
\label{G}
G(\mu^2) = \frac{1}{\mu^2}
  \left[F^{-1}(-\mu^2)\right]^2F'\left[F^{-1}(-\mu^2)\right]
\end{equation}

\section{Specialization to $p$-adic Inflation}
\label{p_compare}

We now consider how our previous results apply to the case of $p$-adic inflation (\ref{p_action}).
For the dimensionful parameters we have
\begin{equation}
  \gamma^4 = \frac{m_s^4}{g_p^2} = \frac{m_s^4}{g_s^2}\frac{p^2}{p-1}
\end{equation}
and $m_s$ is identified with the string mass scale.  Notice that $\gamma > m_s$ for typical
values of $g_s$ and $p$.  The dimensionless coefficients appearing in the potential (\ref{pot})
are
\begin{eqnarray}
  U_0 &=& \frac{1}{2}\frac{p-1}{p+1} \\
  \mu^2 &=& p-1 \\
  g &=& -\frac{p}{2}(p-1)
\end{eqnarray}
The function $F$ associated with the kinetic operator is
\begin{equation}
  F(z) = 1-p^{-z/2}
\end{equation}
so that
\begin{eqnarray}
  F^{-1}(-\mu^2) &=& -2 \label{pFinv} \\
  F'\left[F^{-1}(-\mu^2)\right] &=& \frac{p}{2}\ln p \label{pFp}\\
  G(\mu^2) &=& \frac{2p\ln p}{p-1} \label{pG}
\end{eqnarray}
Notice also that (\ref{gamma_eliminate}) can be used to eliminate the ratio $m_s/M_p$ in 
favour of $p$, $g_s$ as
\[
  \left(\frac{m_s}{M_p}\right)^2 = \frac{8(p+1)}{p^2}\frac{g_s^2}{|n_s-1|}
\]

Using equations (\ref{pFinv}--\ref{pG}) it is
easy to show that we exactly reproduce the expressions for $\lambda$, $\phi_2$, $H_0$, $H_2$ 
and $n_s-1$ given in \cite{pi}.  However, our results for $u_{\mathrm{end}}$ and $A_\zeta^2$
(equation (\ref{u_end}) and (\ref{amp})) differ somewhat from the results of \cite{pi}.
To see this we first compute $u_{\mathrm{end}}$ for $p$-adic inflation using (\ref{u_end})
\begin{equation}
\label{discrepancy}
  \frac{1}{u_{\mathrm{end}}^2} = p\,\left[ \frac{1}{2}\frac{p+1}{p-1}\ln p \right]
\end{equation}
Now, using (\ref{amp}) to compute the COBE normalization for $p$-adic inflation we find
\begin{equation}
\label{amp_discrepancy}
  A_\zeta^2 = \frac{8}{3\pi^2}\frac{g_s^2}{p}
  \frac{e^{N_e|n_s-1|}}{|n_s-1|^3}\,\left[\frac{1}{2}\frac{p+1}{p-1}\ln p\right]
\end{equation}

Equations (\ref{discrepancy}) and (\ref{amp_discrepancy}) should be compared to equations
(5.18) and (5.20) in \cite{pi}.  In both cases the equations differ by a factor of 
$(p+1)\ln p / \left[2(p-1)\right]$.  This
discrepancy arises because the end of inflation was defined here as the end of the validity
of the perturbative expansion is powers of $u$, whereas in \cite{pi} the end of inflation
was defined using the friction-dominated approximation, which has a longer range of validity.
(This discrepancy was noted also in \cite{lidsey}.)
Thus, in the particular case of $p$-adic inflation, the slow-roll dynamics actually 
persist even after the perturbative description (\ref{series_ansatz}) has 
broken down.  
Thus, choosing (\ref{u_end}) to define the end of inflation is more stringent.

Notice that in the case of $p$-adic inflation this discrepancy does not lead to any significant
{quantitative} change in the results since the factor $(p+1)\ln p / \left[2(p-1)\right]$ 
is order unity for the values of $p$ which were considered in \cite{pi}, and which will
also interest us here.

\section{Calculation of the Nongaussianity}
\label{sec:ng}
We now proceed to estimate the level of nongaussianity in nonlocal models, based on
the bispectrum.  As explained above, a full calculation including metric contributions
appears to be prohibitively difficult; we do not know how to quantize the metric
perturbations due to the complexity of the stress-energy tensor.  However using local
scalar field theories as a guide, we hypothesize that the scalar contribution to the
curvature perturbation is of the same order of magnitude as that of the metric contribution, so
that a valid estimate can be obtained from the scalar perturbation alone.  Its
contribution to the bispectrum comes from the cubic term in the action for small
fluctuations.

\subsection{Perturbing the Lagrangian}

It is straightforward to perturb the Lagrangian (\ref{action}) up to
cubic order in the field $\delta \varphi = A \delta \phi$.  The quadratic Lagrangian
is
\begin{equation}
\label{L2}
  \mathcal{L}^{(2)} = \frac{\omega^2}{2\mu^2}
  \left[\delta \varphi \Fbox \delta \varphi + \mu^2 (\delta \varphi)^2\right]
\end{equation}
The cubic Lagrangian is
\begin{equation}
\label{L3}
  \mathcal{L}^{(3)} = c_H (\delta \varphi)^3
\end{equation}
where we have defined
\begin{equation}
\label{cH}
  c_H =  -\frac{g}{3}\frac{m_s^3}{\gamma^2}\left[\frac{F^{-1}(-\mu^2)}{(-\mu^2)}\right]^{3/2}
\end{equation}

\subsection{Calculation of the Bispectrum}

Variation of the quadratic Lagrangian (\ref{L2}) yields the free perturbation equation
of motion (\ref{fluct}) whose solutions are given by (\ref{Fourier}, \ref{xi}, \ref{mode}), 
as we have already discussed.  To leading order in slow-roll parameters the order
of the Hankel functions in (\ref{mode}) is $\nu \cong 3/2$ so that
\begin{equation}
\label{mode2}
  \varphi_{\bf k}(\tau) \cong e^{i\delta} \frac{H_0}{\sqrt{2k^3}} 
  e^{-ik\tau}\left(1+ik\tau\right)
\end{equation}
where we have employed conformal time $\tau$, which is related to cosmic time $t$ by
$a d\tau = dt$.  It follows that $a \cong e^{H_0 t} \cong -1/(H_0\tau)$.  In 
(\ref{mode2}) $k$ is the comoving wave number of the perturbation
and $\delta$ is a constant real phase whose value is irrelevant for our calculation.

The quantity of interest for our calculation is the 3-point correlation function of the curvature
perturbation $\langle\zeta_{\bf k_1}\zeta_{\bf k_2}\zeta_{\bf k_3} \rangle$ where the expectation
value is computed in the interaction vacuum 
$e^{-i\int_{t_0}^{t}dt' H_{\mathrm{int}}(t')}|0\rangle$. The leading contribution is
\begin{eqnarray}
\label{bispectrum}
  \langle\zeta_{\bf k_1}\zeta_{\bf k_2}\zeta_{\bf k_3}(\tau) \rangle
   &=& -i \int_{\tau_i}^{\tau}d\tau'\langle 0 | 
\left[\zeta_{\bf k_1}\zeta_{\bf k_2}\zeta_{\bf k_3}(\tau),
 H_{\mathrm{int}}(\tau')\right] | 0\rangle \\
  &=& -i c_\zeta^3 \int_{\tau_i}^{\tau}d\tau'\langle 0 | 
\left[\xi_{\bf k_1}\xi_{\bf k_2}\xi_{\bf k_3}(\tau),   H_{\mathrm{int}}(\tau')\right] | 0\rangle
\end{eqnarray}
where we have used the fact that
\[
  \zeta_{\bf k}(\tau) = c_{\zeta}\xi_{\bf k}(\tau) = c_{\zeta}\left[
  a_{\bf k} \varphi_{\bf k}(\tau) + a_{\bf -k}^{\dagger} \varphi_{\bf -k}^{\star}(\tau) \right]
\]
The interaction Hamiltonian takes the form
\begin{eqnarray}
  H_{\mathrm{int}} &=& - \int d^{\,3}x\, a^3 \mathcal{L}^{(3)} \\
  &=& -c_H \int d^{\,3}x\, a^3 \delta\varphi^3
\end{eqnarray}
where $c_H$ is defined in (\ref{cH}).

In order to compute (\ref{bispectrum}) we first consider the commutator
\begin{eqnarray}
  \left[\xi_{\bf k}(\tau),\delta\varphi(\tau',{\bf x'})\right]
  &=& \frac{e^{-i{\bf k}\cdot {\bf x'}}}{(2\pi)^{3/2}}
      \left(\varphi_{\bf k}(\tau)\varphi_{\bf k}^{\star}(\tau') - 
      \varphi_{\bf k}(\tau')\varphi_{\bf k}^{\star}(\tau)\right) \nonumber \\
  &\cong& -i \frac{H^2}{3}\left(\tau-\tau'\right)^3
              \frac{e^{-i{\bf k}\cdot {\bf x'}}}{(2\pi)^{3/2}} \label{commutator}
\end{eqnarray}
where on the last line we have taken the large scale limit $-k\tau, -k\tau' \ll 1$.

Noting that the commutator (\ref{commutator}) is a c-number we have
\[
  \left[\xi_{\bf k}(\tau),\delta\varphi^3(\tau',{\bf x'})\right]
  = 3 \left[\xi_{\bf k}(\tau),\delta\varphi(\tau',{\bf x'})\right] \delta \varphi^2(\tau',{\bf x'})
\]
The expectation value in (\ref{bispectrum}) can be written as
\begin{eqnarray}
  \langle 0 | \left[\xi_{\bf k_1}\xi_{\bf k_2}\xi_{\bf k_3}(\tau),
  \delta\varphi^3(\tau',{\bf x'}) \right]|0\rangle &=& 
  3 \left[\xi_{\bf k_3}(\tau),\delta\varphi(\tau',{\bf x'})\right] 
  \langle 0 | \xi_{\bf k_1}\xi_{\bf k_2}(\tau) \delta\varphi^2(\tau',{\bf x'})|0\rangle \nonumber \\
+ && \hspace{-2mm}
3 \left[\xi_{\bf k_2}(\tau),\delta\varphi(\tau',{\bf x'})\right] 
  \langle 0 | \xi_{\bf k_1}\delta\varphi^2(\tau',{\bf x'})\xi_{\bf k_3}(\tau)|0\rangle \nonumber\\
+ && \hspace{-2mm}
3 \left[\xi_{\bf k_1}(\tau),\delta\varphi(\tau',{\bf x'})\right] 
  \langle 0 | \delta\varphi^2(\tau',{\bf x'})\xi_{\bf k_2}\xi_{\bf k_3}(\tau)|0\rangle
\label{contractions}
\end{eqnarray}
Consider, for example, the first term.  Carrying out the Wick contractions we have
\begin{eqnarray*}
 && 3 \left[\xi_{\bf k_3}(\tau),\delta\varphi(\tau',{\bf x'})\right] 
  \langle 0 | \xi_{\bf k_1}\xi_{\bf k_2}(\tau) \delta\varphi^2(\tau',{\bf x'})|0\rangle \\
 &=& 6 \left[\xi_{\bf k_3}(\tau),\delta\varphi(\tau',{\bf x'})\right] 
  e^{-i({\bf k_1} + {\bf k_2})\cdot {\bf x'}} D_{\bf k_1}(\tau-\tau')D_{\bf k_2}(\tau-\tau')
\end{eqnarray*}
where the momentum space propagator is
\begin{eqnarray}
  D_{\bf k}(\tau-\tau') 
  &=& \frac{1}{(2\pi)^{3/2}} \varphi_{\bf k}(\tau) \varphi^{\star}_{\bf k}(\tau')\nonumber \\ 
  &=& \frac{1}{(2\pi)^{3/2}} \frac{H^2}{2k^3}e^{-ik(\tau-\tau')}(1+ik\tau)(1-ik\tau') \nonumber \\
  &\cong& \frac{1}{(2\pi)^{3/2}} \frac{H^2}{2k^3} \label{propagator}
\end{eqnarray}
and on the last line we have restricted to large scales.  The remaining terms in 
(\ref{contractions}) are similar.  

We are now in a position to calculate (\ref{bispectrum}).  On large scales 
$-k_i\tau, -k_i\tau' \ll 1$ we have
\begin{equation}
\label{intermediate}
  \langle\zeta_{\bf k_1}\zeta_{\bf k_2}\zeta_{\bf k_3}(\tau) \rangle
   \cong - c_H c_\zeta^3\frac{H^2}{2} \int_{\tau_i}^{\tau}d\tau' \frac{(\tau-\tau')^3}{(\tau')^4}
   \frac{k_1^3 + k_2^3 + k_3^3}{k_1^3k_2^3k_3^3} 
   \delta^{(3)}\left({\bf k_1} +{\bf k_2}+{\bf k_3} \right)
\end{equation}
The time integral is divergent and needs to be regulated.  
Taking $\tau$ to correspond to the end of inflation and $\tau_i$ 
to correspond to the beginning of inflation we have
\begin{equation}
\label{cutoff}
  \int_{\tau_i}^{\tau}d\tau' \frac{(\tau-\tau')^3}{(\tau')^4} 
  \cong \ln \left(\frac{\tau_i}{\tau}\right) = \ln e^{N_e} = N_e
\end{equation}
where $N_e$ is the total number of e-foldings of inflation.

It is worth commenting on the infrared (IR) divergence in the $d\tau$ integral in 
(\ref{intermediate}) which does not appear in the standard calculation of $f_{NL}$ 
\cite{Maldacena,large_nongauss}.  Naively, this divergence would seem to suggest that the 
curvature perturbation in $\zeta$ is not freezing out on super-horizon scales, as it does in 
the standard theory.  However, we suspect that the IR divergence in (\ref{cutoff}) is in fact 
an artifact of the approximations which we have made.  Indeed, a calculation of $f_{NL}$ in 
a local field theory which makes precisely the same approximations as we have made 
(neglecting the metric perturbations and the departure of the background from pure de Sitter 
space) yields exactly the same IR divergence \cite{div}.  It is clear from the 
calculation of \cite{div} that this IR divergence will generically arise in the three-point 
function of any light scalar field in de Sitter space.  In appendix B we show that,
despite the presence of the IR divergence, the calculation of \cite{div} can be used to
estimate $f_{NL}$ in a local field theory and their result does reproduce the result of the 
complete calculation \cite{Maldacena} up to factors of order unity.  We assume that
the same holds true in the nonlocal theories which we consider.

Though we believe that in a more complete calculation one would find that $\zeta$ is conserved
outside the horizon in the nonlocal theory, we were not able to demonstrate this conclusively.
A proper resolution of this issue would require us to perform a complete second
order cosmological perturbation calculation, including metric perturbations and departures of
the background from pure de Sitter space.  Such a calculation is complicated
by the nonlocal structure of the theory and is beyond the scope of the present work.
Though it is a logical possibility that $\zeta$ is actually not conserved in the nonlocal
theory, it would be surprising.
Quite generally, the time derivative of the large scale curvature perturbation is proportional
to the nonadiabatic pressure, $\dot{\zeta} \propto P_{\mathrm{nad}}$ \cite{conserved}.
In a model with only one degree of freedom there are no entropy perturbations and
$P_{\mathrm{nad}} = 0$ so that $\dot{\zeta} = 0$ on large scales.  Thus we expect that
as long as (\ref{action}) describes only a single degree of freedom then the large
scale curvature perturbation should be conserved (in a complete calculation).  It is well
known that higher derivative theories with \emph{finitely} many derivatives can be rewritten
in terms of multiple scalar fields (at least one of which will generically be ghost-like).
However, for \emph{infinite} derivative theories of the type which we consider this need not be
the case \cite{tirtho}.  For infinite derivative theories one should consider the pole structure
of the propagator $G_k \sim \left[F(-k^2/m_s^2) + \mu^2\right]^{-1}$.  Our restriction to models 
where $F^{-1}(-\mu^2)$ is single-valued (below equation (\ref{omega}))  ensures that the 
propagator has only one pole (at least near the false vacuum $\phi = 0$).\footnote{For example, 
in the case of $p$-adic inflation the propagator near the unstable maximum has
only one pole corresponding to a scalar field with mass-squared $-2m_s^2$: the bosonic
open string tachyon.  In contrast there are \emph{no} poles about the true minimum,
which is the $p$-adic version of the statement that there are no open strings in the
tachyon vacuum.}  It follows that our analysis is restricted to infinite derivative theories 
which describe only a single scalar degree of freedom and hence one expects 
$P_{\mathrm{nad}} = 0$ (and therefore $\dot{\zeta}\cong 0$) for such theories.

\subsection{The Nonlinearity Parameter}

We now estimate the nonlinearity parameter by comparing (\ref{intermediate})
to the WMAP ansatz
\begin{equation}
\label{WMAP_ansatz}
   \langle\zeta_{\bf k_1}\zeta_{\bf k_2}\zeta_{\bf k_3}(\tau) \rangle
   = (2\pi)^7 \left(-\frac{3}{10}f_{NL}\right) A_\zeta^4
   \frac{k_1^3 + k_2^3 + k_3^3}{k_1^3k_2^3k_3^3} 
   \delta^{(3)}\left({\bf k_1} +{\bf k_2}+{\bf k_3} \right)
\end{equation}
where, as we have shown previously, 
$A_\zeta^2 = H^4/ (2\pi \dot{\varphi}_0)^2 = c_{\zeta}^2 H_0^2 / (4\pi^2)$.  
Comparing (\ref{WMAP_ansatz}) to (\ref{intermediate}) we have
\begin{equation}
\label{f_NL_1}
  f_{NL} = \frac{5 N_e}{3(2\pi)^5} c_H c_\zeta \frac{1}{A_\zeta^2}
\end{equation}
It is straightforward to compute the product $c_H c_\zeta$ using (\ref{cH}) and 
(\ref{prefactor}).  The result is
\begin{equation}
\label{cHczeta1}
  c_H c_\zeta = -\frac{g}{3 U_0^{1/2}}\left(\frac{m_s}{\gamma}\right)^4
                 \frac{e^{\frac{N_e}{2}|n_s-1|}}{|n_s-1|}
                 \left(\frac{F^{-1}(-\mu^2)}{(-\mu^2)}\right)^{3/2}
                 \sqrt{G(\mu^2)}
\end{equation}
where $G(\mu^2)$ is defined in (\ref{G}).  It is possible to further simplify this expression
by using (\ref{gamma_eliminate}) to eliminate $\gamma^{-4}$ with the result
\begin{equation}
\label{cHczeta2}
  c_H c_\zeta = -\frac{g U_0^{1/2}}{6}\left(\frac{m_s}{M_p}\right)^2
                 e^{\frac{N_e}{2}|n_s-1|} \frac{|F^{-1}(-\mu^2)|^{1/2}}{|\mu|^3}\sqrt{G(\mu^2)}
\end{equation}
We can eliminate $(m_s / M_p)^2$ from (\ref{cHczeta2}) using the COBE normalization
(\ref{amp}).  The final result is
\begin{equation}
\label{f_NL}
  f_{NL} \cong -\frac{5 N_e}{48\pi^3} g U_0^{1/2} e^{-\frac{N_e}{2}|n_s-1|} |n_s-1|^2 \, K(\mu^2)
\end{equation}
where we have defined
\begin{equation}
\label{K}
  K(\mu^2) = \frac{1}{\mu^2}\frac{1}{|F^{-1}(-\mu^2)|^{1/2}}
           \frac{1}{|F'\left[F^{-1}(-\mu^2)\right]|^{1/2}}
\end{equation}
Equation (\ref{f_NL}) is the main result of this section.

\subsection{The Perturbative Regime}
\label{sec:pert}

Since $f_{NL} \propto g$ and we are interested in models with $|f_{NL}| \gg 1$ it is important
to understand what values of $g$ are acceptable.  In conventional (local) inflationary models
one would require $|g| \ll 1$ in order not to spoil slow roll; however, this need not be the
case for the nonlocal hill-top models which we consider.  Indeed, in the case of $p$-adic
inflation one has $|g| \sim p^2 \gg 1$ (see section \ref{p_compare}).  In terms of the canonical
field the Lagrangian is of the form 
\begin{equation}
\label{can_lag}
  \mathcal{L} 
  = \frac{\omega^2}{2\mu^2} \varphi\, F\left(\frac{\Box}{m_s^2}\right)\varphi
   - V(\varphi)
\end{equation}
where
\begin{equation}
\label{can_pot}
  V(\varphi) = V_0 - \frac{\omega^2}{2}\varphi^2 + c_H \,\varphi^3
\end{equation}
and $V_0 = \gamma^4 U_0$.  In a local field theory the one-loop correction to the cubic
term of the potential (\ref{can_pot}) would be of the order
\[ 
  \delta c_H \sim \frac{1}{16\pi^2}\frac{c_H^3}{\omega^3}
\]
so that the dimensionless quantity $c_H^2 / \omega^2$ controls the perturbative expansion.
We therefore expect that the calculation is under control as long as $c_H^2 < \omega^2$.
Using equations (\ref{cH}) and (\ref{omega}) the condition $c_H^2 < \omega^2$
implies the following upper bound on $g$:
\begin{equation}
\label{g_bound}
  |g| < \frac{3\mu^3}{|F^{-1}(-\mu^2)|}\left(\frac{\gamma}{m_s}\right)^{2}
\end{equation}
In a given model the largest reliable value of $f_{NL}$ is achieved by choosing $g$
to saturate the bound (\ref{g_bound}).

\section{Examples}
\label{sec:png}

\subsection{$p$-adic Inflation}

With the results of section \ref{p_compare} we can compute $f_{NL}$ 
for the $p$-adic inflation model, using (\ref{f_NL}, \ref{K}).  The result is
\[
  f_{NL} \cong \frac{5 N_e}{192\pi^3} e^{-\frac{N_e}{2}|n_s-1|} |n_s-1|^2 \sqrt{p}\, 
               \left[ \frac{2}{\ln p}\frac{p-1}{p+1} \right]^{1/2}
\]
so that $f_{NL} \sim \sqrt{p}$ for $p\gg 1$.   
This result may slightly underestimate the nongaussianity 
produced in $p$-adic inflation because of the extra factor 
$2(p-1)/\left[(p+1)\ln p\right]$,  
which appears because the perturbative approach breaks down before
 the apparent end of the slow roll phase.
This is precisely the discrepancy discussed below (\ref{amp_discrepancy}).  Taking into account
the fact that in $p$-adic inflation slow roll ends when $u \sim p^{-1/2}$, rather than
when $u$ is given by (\ref{discrepancy}), we obtain
\begin{equation}
\label{final1}
  f_{NL} \cong  \frac{5 N_e}{192\pi^3} e^{-\frac{N_e}{2}|n_s-1|} |n_s-1|^2 \, \sqrt{p}
\end{equation}
For $n_s \cong 0.95$ and $N_e \cong 60$ this implies
\begin{equation}
\label{final2}
  f_{NL} \cong 2.8\times 10^{-5} \, \sqrt{p}
\end{equation}
so that large $f_{NL}$ requires large values of $p$.

How large can (\ref{final2}) be made?  
The COBE normalization relates $g_s$ and $p$ as \cite{pi}
\[
  g_s = \frac{5\pi\sqrt{3}}{2\sqrt{2}}10^{-5}e^{-\frac{N_e}{2}|n_s-1|}|n_s-1|^{3/2}\sqrt{p}
\]
For $N_e\cong 60$ and $n_s\cong 0.05$ the condition $g_s < 1$ bounds $p$ from above
as 
\[
  p \lsim 1.7\times 10^{13}
\]
so that, using (\ref{final2}), the nonlinearity parameter is bounded as
\[
  f_{NL} \lsim 120
\]
with the upper limit corresponding to $g_s = 1$.  The maximum possible nongaussianity
is within the observational limit $f_{NL} < 300$ though it should be observable in future
missions. We see that large nongaussianity is possible in $p$-adic
inflation.  It is interesting that the largest values of $f_{NL}$ correspond to $g_s$
close to unity, which is considered natural from the string theory perspective.

In evaluating out estimate for $f_{NL}$, eq.\ (\ref{final2}), we have used
the total number of e-foldings between horizon crossing and the end of inflation 
$N_e \cong 60$, rather than the number of e-foldings which can be observed.  
This is justified because
the factor $N_e$ has its origins in an IR divergence (\ref{cutoff}).  This divergence should
be regulated by the largest scale to which we have experimental access, namely, our current 
horizon size.  The need for such IR regulators in cosmological perturbation theory
was discussed in \cite{L&B} where it was argued that these divergences simply reflect
 our ignorance about scales beyond the horizon.
Notice that typically the total number of e-foldings between horizon crossing and the end of 
inflation is $N_e \sim |n_s-1|^{-1}$ so that one of the factors of $n_s-1$ in 
(\ref{f_NL}) is cancelled by the factor $N_e$.  An identical cancellation occurs in local field 
theory models, see appendix B. 

Finally, we note that even taking $N_e \cong 6$ we would still obtain $f_{NL} \sim 12 g_s$
which, for $g_s \sim 1$, is still several orders of magnitude larger than the prediction of
the simplest inflationary models.  As long as $f_{NL} > \mathcal{O}(1)$ then the primordial
nongaussianity should be detectable in future missions \cite{future}.  For the
primordial $f_{NL} = \mathcal{O}(1)$ the situation is more complicated since post-inflationary
super-horizon evolution will also generate an order unity contribution to the nonlinearity 
parameter \cite{post_inf_NG} (see also \cite{post_inf_NG2}) and it may be difficult to 
disentangle these two distinct sources of nongaussianity.

In principle one could make $f_{NL}$ even larger than $f_{NL} \sim 120$ 
in this model by relaxing the requirement  that $g_s < 1$, however, at large 
$g_s$ we can no longer rely upon
perturbation theory to compute the bispectrum or other quantities from the tree-level
$p$-adic action.  String loop effects will invalidate this action when $g_s\gg 1$.
This is also consistent
with our discussion in subsection  \ref{sec:pert} since the bound (\ref{g_bound}), for
the case of $p$-adic inflation, corresponds  to $g_s < 3$.

\subsection{A Toy Model with Exponential Kinetic Function}

As an example of further possibilities within the framework of nonlocal field theories,
we next investigate a toy model which can give rise to significant nongaussianity.  We choose
a kinetic function
\begin{equation}
\label{toyF}
  F(z) = \frac{1}{\alpha}\left( 1 - e^{-\beta z}\right)
\end{equation}
where it is assumed that $\alpha,\beta > 0$.  Equation (\ref{toyF}) is typical of the kinds of
kinetic functions which arise in string field theory.  It is straightforward to compute
\begin{eqnarray}
  F^{-1}(-\mu^2) &=& -\frac{1}{\beta}\ln\left(1+\alpha \mu^2\right) \label{toy_f_inv}\\
  F'\left[F^{-1}(-\mu^2)\right] &=& \frac{\beta}{\alpha}\left(1+\alpha \mu^2\right)
\end{eqnarray}
so that
\begin{eqnarray}
  G(\mu^2) &=& \frac{1}{\beta}\left(\frac{1+\alpha\mu^2}{\alpha\mu^2}\right)
  \ln^2\left(1+\alpha\mu^2\right) \\
  K(\mu^2) &=& \frac{1}{\mu^3}\sqrt{\frac{\alpha\mu^2}{1+\alpha\mu^2}}\,
  \sqrt{\frac{1}{\ln\left(1+\alpha\mu^2\right)}}
\end{eqnarray}
In passing, it is interesting to compute the effective mass of the field fluctuation for this
model using (\ref{toy_f_inv}) and (\ref{omega})
\[
  \omega^2 = \frac{m_s^2}{\beta}\ln\left(1+ \alpha\mu^2\right)
\]
The fluctuations $\delta\varphi$ behave as though the field had mass-squared $-\omega^2$.  
However, for an exponential kinetic function (\ref{toyF}) this effective mass is only
logarithmically sensitive to the actual mass $\mu$.  This explains the novel behaviour
first noted in \cite{pi} that the cosmology of the $p$-adic tachyon is virtually insensitive
to the naive mass of the field.

We set
\begin{equation}
\label{toygamma}
  \gamma^4 = \frac{m_s^4}{g_s^2}
\end{equation}
by analogy with the D-brane tension.  Taking $U_0 = 1$ the equation (\ref{gamma_eliminate})
fixes the string scale as
\begin{equation}
\label{toym_s}
  \left(\frac{m_s}{M_p}\right)^2 = \frac{2 g_s^2}{|n_s-1|}\frac{\ln(1+\alpha\mu^2)}{\beta}
\end{equation}
while the COBE normalization requires
\begin{equation}
\label{toyCOBE}
  \frac{g_s^2}{\beta^2} = 6\pi^2 |n_s-1|^3e^{-N_e|n_s-1|}\frac{\alpha\mu^2}{1+\alpha\mu^2}
                          \frac{1}{\ln^3\left(1+\alpha\mu^2\right)} \cdot 25 \cdot 10^{-10}
\end{equation}
If we take $\alpha = \mu^2 = 1$ then (\ref{toyCOBE}) requires
\begin{equation}
\label{beta}
  \frac{\beta}{g_s} = 10^6
\end{equation}
and (\ref{toym_s}) gives $(m_s / M_p)^2 = 3\times 10^{-5}\, g_s$ so that $m_s \ll M_p$
unless $g_s$ is very large.  

To obtain an upper bound on $f_{NL}$ we
note that $g$ is bounded from above by 
(\ref{g_bound}) so that
\begin{equation}
\label{toy_g}
  g < \frac{3\mu^3}{\ln(1+\alpha\mu^2)}\frac{\beta}{g_s} = \frac{3}{\ln 2}\frac{\beta}{g_s}
      = 0.4\times 10^{7}
\end{equation}
(we are taking $g>0$).
Using (\ref{beta}), (\ref{toy_g}), $n_s \cong 0.95$ and $N_e \cong 60$ we obtain
\begin{equation}
\label{toy_f_NL}
  |f_{NL}| \cong 9.5\times 10^{-5}\, g
\end{equation}
so that $f_{NL} \lsim 300$ bounds $g$ from above as $g \lsim 0.31\times 10^7$.  Since
this upper bound is stronger than the bound (\ref{toy_g}) it follows that it is possible 
to saturate the observational limit on $f_{NL}$.

\section{Conclusions}

In this paper we have studied the nongaussianity produced during nonlocal hill-top
inflation. We were particularly interested in $p$-adic inflation, in which case large
nongaussianity is possible, and indeed is natural since the upper bound on $f_{NL}$
corresponds to  $g_s \sim 1$.  We also considered a toy model with an exponential
kinetic function, typical of string field theory Lagrangians, and showed that for
certain values of the parameters large nongaussianity is possible in this model. 
Thus, nonlocal hill-top inflation models are among the few inflationary scenarios which
can give rise to $f_{NL} \gg 1$.

There are several caveats to our work.  In computing the curvature perturbation $\zeta$ we
have  neglected both metric perturbations and also departures of the background from
pure de Sitter space.  Although these approximations reproduce the correct answer up to
acceptable accuracy in standard (local) theories, it is not clear if this is also
true in the nonlocal theories which we have considered.  A cause for concern is the IR
divergence (\ref{cutoff}) which we have regulated by the number of e-foldings of
inflation.  This divergence does not occur in the standard theory and seems to indicate
that our definition of $\zeta$ does not freeze out on large scales, as it should.   We
speculated that this divergence is an artifact of the approximations which we 
made, and this claim is supported by the fact that an identical divergence occurs also
in the local theory when one makes exactly the same approximations which we have
made \cite{div}.  That being said, we have not proven that $\zeta$
will really freeze out in a more comprehensive treatment.  It is possible that the relation
$\zeta \sim -H\delta\varphi / \dot{\varphi}_0$ does not hold for nonlocal theories.  In
order to answer this question in a satisfactory manner it will be necessary to perform
a complete and rigorous  analysis of the cosmological perturbations in this theory,
incorporating also  metric perturbations.  Such an analysis is complicated by the fact
that the kinetic operator $F(\Box)$ will be of the form $F(\Box_0 + \delta \Box)$ where
$\Box_0$ is the covariant d'Alembertian associated with the homogeneous background 
and $\delta
\Box$ is the perturbation which contains both metric perturbations $\delta g_{\mu\nu}$
and also derivatives.  In general $\Box_0$ and $\delta\Box$ do not commute, making
it extremely difficult to solve the nonlocal evolution equations.  We leave a
rigorous treatment of the cosmological perturbations to future analysis.  

In a local field theory the gravitational sector is known to give a small contribution to
the nongaussianity and hence the neglect of the metric perturbations is justified when
$f_{NL} \gg 1$ \cite{large_nongauss}.  It is not clear if this
is also the case in nonlocal theories.  In a complete calculation one would find terms 
in the perturbed Lagrangian which involve the operator $F(\Box_0 / m_s^2)$ acting on the 
scalar metric perturbations and hence it is not clear that 
the contribution of such terms to $f_{NL}$ will be suppressed.  Barring anomalous cancellations
we would expect such terms to be of the same order of magnitude as those which we consider
and hence our estimate should be reliable up to factors of order unity.

Another caveat is the complication in setting up the initial value problem for differential
equations with infinitely many derivatives.  It is known that the initial value problem
for equations with infinitely many derivatives is fundamentally different from the
initial value problem for an equation with $N$ derivatives where $N \gg 1$ \cite{zweibach}.
Infinite derivative equations have been studied from a mathematical physics perspective
in \cite{math}.  For further discussions about the initial value problem see 
\cite{pi,justin,caustics}.  

A related difficultly is the issue of quantizing a nonlocal theory.  In our analysis we have 
assumed the usual Bunch-Davies vacuum for the perturbation $\delta\varphi$, though it is not 
clear that this is correct in the present context.  Quantization of nonlocal theories with 
finitely many derivatives generically leads to ghost excitations (and also to a classical 
pathology called the Ostrogradski instability) though nonlocal theories with infinitely many 
derivatives (such as we consider) can evade these difficulties. For further discussion see 
\cite{pi,tirtho,justin}.

Despite these difficulties, we believe that our calculation does provide a rough (order of 
magnitude) estimate of the actual nongaussianity produced.  It is easy to intuitively understand
why $f_{NL}$ can be made large in these kinds of models.  Our result (\ref{f_NL}) is equivalent
to computing the three-point function for a light scalar field in de Sitter space.  This 
correlator is proportional to coefficient of the cubic term in the potential, $g$, and hence 
will be large when $g$ is large.  In conventional inflationary models this is impossible since 
a large coupling $g$ would spoil slow roll.  However, in nonlocal theories this is not 
necessarily true. (Similarly a large mass term $\mu^2$ would spoil inflation in a local theory, 
though there is no problem with exponentially large $\mu^2$ in our models.)  Indeed, in the 
examples where we have found large nongaussianity this is simply because $g \gg 1$, 
and the novelty is that such theories can support inflation.

\section*{Acknowledgments}

This work was supported in part by NSERC and FQRNT.  
We are grateful to X.\ Chen for many helpful discussions, comments 
and for collaboration during the early stages of this work.
Thanks also to R.\ Brandenberger, D.\ Ghoshal, J.\ Lidsey 
and E.\ Lim for enlightening discussions.

\renewcommand{\theequation}{A-\arabic{equation}}
\setcounter{equation}{0}  

\section*{APPENDIX A: $p$-adic Scalar Field Evolution Beyond Linear Order}

In this appendix we demonstrate that the equivalence between local and nonlocal theories
breaks down beyond linear order in perturbation theory by studying the $p$-adic scalar 
field equation of motion
\begin{equation}
\label{pKG}
  p^{-\Box / 2} \psi = \psi^p
\end{equation}
(in units $m_s \equiv 1$, which we employ throughuot this appendix) up to second order in 
perturbation theory.  For simplicity we neglect metric perturbations and assume pure de Sitter 
background.  Expanding the $p$-adic scalar in perturbation theory as
\begin{eqnarray*}
  \psi &=& 1 + \phi \\
       &=& 1 + \done\phi + \frac{1}{2}\dtwo\phi
\end{eqnarray*}
It is straightforward perturb the field equation (\ref{pKG}) up to second order with the result
\begin{eqnarray}
  \left[ p^{-\Box/2} - p \right] \done \phi &=& 0 \label{lin_eom}\\
  \left[ p^{-\Box/2} - p \right] \dtwo \phi &=& p(p-1)\left(\done \phi \right)^2
  \label{sec_eom}
\end{eqnarray}
At linear order one may construct solutions of (\ref{lin_eom}) by taking 
$\Box \done \phi = -\omega^2 \done \phi$, however, this prescription fails for the second
order equation (\ref{sec_eom}).  It is interesting to notice that in terms of 
the canonical field 
\[
  \varphi = \frac{p}{\sqrt{2}\, g_s} \phi
\]
the second order equation (\ref{sec_eom}) becomes
\begin{equation}
\label{kg2}
\left[ p^{-\Box/2} - p \right] \dtwo \varphi = c\left(\done \varphi \right)^2
\end{equation}
where
\[
  c = \frac{p(p-1)}{A} = \sqrt{2} g_s (p-1)
\]
The COBE normalization gives $g_s \sim 10^{-7} \sqrt{p}$ (for $n_s\cong 0.95$ 
and $N_e \cong 60$) so that $c$ is of the order
\begin{equation}
\label{c}
  c \sim 10^{-7} p^{3/2}
\end{equation}
for $p\gg 1$.
We see that the second order effects can be made large by taking $p \gg 1$ which is precisely
the regime in which $f_{NL} > 1$ occurs.  As one might expect, this also coincides
with the regime in which the nonlocal structure of the theory is playing an important
role in the dynamics, as emphasized in \cite{pi}.

\renewcommand{\theequation}{B-\arabic{equation}}
\setcounter{equation}{0}  

\section*{APPENDIX B: Comparison to Local Theory}

In \cite{div} Falk {\it et al}.\ studied the three-point function of the inflaton perturbation
making the same approximations as we have made in our calculation.  Namely, \cite{div} neglect
metric perturbations and departures of the background expansion from pure de Sitter.  For
the potential
\begin{equation}
\label{pot_app}
  V(\varphi) = V_0 - \frac{g}{6}\varphi^3
\end{equation}
Falk {\it et al}.\ find
\begin{equation}
\label{bi_app}
  \langle\delta\varphi_{k_1}(\tau)\delta\varphi_{k_2}(\tau)\delta\varphi_{k_3}(\tau)\rangle
  \cong +\frac{2\pi^3}{3}g H^2 N_e \frac{\sum_{i=1}^{3}k_i^3}{\Pi_{i=1}^{3}k_i^3}
  \delta^{(3)}\left(\sum_{i=1}^{3}{\bf k}_i\right)
\end{equation}
The factor of $N_e$ in (\ref{bi_app}) arises from regulating a time dependence of
the form $\ln \tau$, exactly as in (\ref{cutoff}).  We now use this result to estimate 
$f_{NL}$ for the potential (\ref{pot_app}), showing that
the answer agrees with a more careful calculation \cite{Maldacena} up to factors 
of order unity.  For simplicity we assume that $g>0$ throughout this appendix and restrict
ourselves to the case where $\varphi$ rolls from the unstable point $\varphi = 0$ towards
some positive value $\varphi>0$.

We assume that $V_0\gg g \varphi^3$ throughout inflation so that $H \cong H_0 
= \sqrt{V_0}/\sqrt{3} M_p$.  The slow roll parameters, $\epsilon = 2^{-1} M_p^2 (V' / V)^2$
and $\eta = M_p^2 V'' / V$, evaluated at the time of horizon crossing $t=t_\star$, are
\begin{eqnarray*}
  \eta &\cong& -M_p^2 \frac{g\varphi_\star}{V_0} \cong -\frac{g\varphi_\star}{3H_0^2} \\
  \epsilon &\cong& |\eta|\frac{g \varphi_\star^3}{V_0} \ll |\eta|
\end{eqnarray*}
where $\varphi_\star = \varphi(t=t_\star)$ is the value of the inflaton at horizon
crossing.  The spectral tilt is
\begin{equation}
\label{app_n}
  n_s - 1 = 2\eta -6\epsilon \cong 2\eta
\end{equation}
The slow roll Klein-Gordon equation
\begin{equation}
\label{app_kg}
  3 H \dot{\varphi} \cong -V'
\end{equation}
has solution
\begin{equation}
\label{app_soln}
  \varphi(N) \cong \varphi_\star\left[1-\frac{|\eta|}{2}N\right]^{-1}
\end{equation}
where $N = H_0(t-t_\star)$ so that $\varphi(N=0) = \varphi_\star$ is the inflaton 
value at horizon crossing.  It is straightforward to see that
\[
  \frac{1}{2M_p^2}\frac{\dot{\varphi}^2}{H^2} 
  \cong \frac{\eta^2 \varphi_*^2}{8M_p^2}\left(1-\frac{|\eta|}{2}N\right)^{-4}
\]
so that inflation ends $N_e$ e-foldings after horizon crossing when $|\eta| N_e / 2 \sim 1$.
It follows that $|n_s-1| N_e = \mathcal{O}(1)$.

We assume that 
\begin{equation}
\label{app_zeta}
  \zeta = c_\zeta \delta \varphi = -\frac{H}{\dot{\varphi}}\delta\varphi
\end{equation}
The value of $c_\zeta$ at horzion crossing is
\begin{equation}
\label{app_c}
  c_\zeta = -\left.\frac{H}{\dot{\varphi}}\right|_{N=0} 
  \cong -\left.\frac{1}{d\varphi/dN}\right|_{N=0} 
  \cong  -\frac{2}{\eta\varphi_\star}
\end{equation}
where in the last equality we have used (\ref{app_soln}).  The COBE normalization is
\begin{equation}
\label{app_COBE}
  A_\zeta^2 = c_\zeta^2 \left(\frac{H_0}{2\pi}\right)^2 = 25\times 10^{-10}
\end{equation}
so that, using (\ref{app_c}), we have
\begin{equation}
\label{app_COBE2}
  A_\zeta^2 \cong \frac{H_0^2}{\pi^2\eta^2\varphi_\star^2}
\end{equation}

Using (\ref{app_zeta}) and 
(\ref{app_c}) we can convert (\ref{bi_app}) into an estimate for the bispectrum of the 
curvature perturbation
\begin{equation}
\label{bi2_app}
  \langle\zeta_{k_1}(\tau)\zeta_{k_2}(\tau)\zeta_{k_3}(\tau)\rangle
  \cong +\frac{2(2\pi)^3}{3} \frac{g H_0^2 N_e}{\eta^3\varphi_\star^3}
  \frac{\sum_{i=1}^{3}k_i^3}{\Pi_{i=1}^{3}k_i^3}
  \delta^{(3)}\left(\sum_{i=1}^{3}{\bf k}_i\right)
\end{equation}
Comparing (\ref{bi2_app}) to the WMAP ansatz (\ref{WMAP_ansatz}) we obtain the following 
estimate for the nonlinearity parameter
\begin{equation}
\label{app_f_NL}
  |f_{NL}| \sim \frac{5}{24} (n_s-1)^2 N_e
\end{equation}
whereas the complete calculation \cite{Maldacena} gives $|f_{NL}| \sim \frac{5}{12}(n_s-1)$.  
The estimate (\ref{app_f_NL}) reproduces the result of the complete calculation up to an order 
of magnitude since $|n_s-1| N_e =\mathcal{O}(1)$.




\bibliographystyle{apsrmp}
\bibliography{rmp-sample}

\end{document}